\documentstyle[preprint,epsfig,eqsecnum,aps]{revtex}

\def\be{\begin{equation}}
\def\ee{\end{equation}}
\def\bea{\begin{eqnarray}}
\def\eea{\end{eqnarray}}

\begin{document}

\title{Quasiparticle Random Phase Approximation with inclusion of
the Pauli Exclusion Principle}

\author{ 
F. \v Simkovic$^1$\thanks{{\it On leave from:}
 Department of Nuclear physics, Comenius University, 
SK--842 15 Bratislava, Slovakia, e-mail: simkovic@fmph.uniba.sk},
A. A. Raduta$^1$\thanks{{\it On leave from:}
Institute of Physics and Nuclear Engineering,
Bucharest, POB MG6, Romania and Dept. of Theoretical Physics and
Mathematics, Faculty of Physics, Bucharest University, POB MG 11,
Romania, 
e-mail: raduta@theor1.theory.nipne.ro },
M. Veselsk\'y$^2$\thanks{{\it On leave from:}
Institute of Physics of Slovak Academy of Sciences,
SK--842 28 Bratislava, Slovakia,
e-mail: fyzimarv@savba.sk },
and Amand Faessler$^1$\thanks{e-mail: amand.faessler@uni-tuebingen.de},
}

\address{
1.  Institute of Theoretical Physics, University of Tuebingen,
D--720 76 Tuebingen, Germany\\
2. Cyclotron Institute, Texas A\&M University, College Station,
TX-77843, USA}

\date{\today}
\maketitle

%%%%%%%%%%%%%%%%%%%%%%%%%%%%%%%%%%%%%%%%%%%%%%%%%%%%%%%%%%%%%%%%%%%%%%%
%                          ABSTRACT                                   %
%%%%%%%%%%%%%%%%%%%%%%%%%%%%%%%%%%%%%%%%%%%%%%%%%%%%%%%%%%%%%%%%%%%%%%%

\begin{abstract}
Limitations of the Quasiparticle Random Phase
Approximation (QRPA) are studied within an exactly solvable model, 
with a two body interaction of Fermi type. A special
attention is paid to the violation of the Pauli
exclusion principle (PEP) in solving the QRPA equation. A comparison of 
the exact solution, obtained by the diagonalization of 
a schematic nuclear Hamiltonian and those obtained 
within the standard QRPA,
the renormalized QRPA, the QRPA with pertubative treatment of the
PEP and the QRPA with exact consideration of the PEP, is presented.
The agreement quality is judged in terms of
the quasiparticle number operator matrix elements in the
ground state and in the first excited states, 
of the $\beta$ transition amplitudes, of the Ikeda sum rule and of the
nuclear matrix element for the double beta decay.  
We have found that restoring the PEP,  
the QRPA solutions are considerably stabilized and
a better agreement with the exact solution is obtained. 
\end{abstract}

%%%%%%%%%%%%%%%%%%%%%%%%%%%%%%%%%%%%%%%%%%%%%%%%%%%%%%%%%%%%%%%%%%%%%%%
%                          INTRODUCTION                               %
%%%%%%%%%%%%%%%%%%%%%%%%%%%%%%%%%%%%%%%%%%%%%%%%%%%%%%%%%%%%%%%%%%%%%%%
\section{Introduction}
\label{sec:level1}

The Quasiparticle Random Phase Approximation (QRPA) has been 
found to be a powerful method for describing many-body systems.
Due to its simplicity, 
the proton-neutron QRPA is the nuclear structure method which has been  
most frequently used to interpret some nuclear structure
aspects of the beta ($\beta$) and double beta ($\beta\beta$)
 decay for open shell systems 
\cite{vog86,eng88,civ91,mut89,tai88,RFSK91,cheon,krm90,krm93,PSV96}.
The QRPA provides a description of excited states
by including some nucleon-nucleon
correlations in the ground state.

The QRPA equations are derived directly from the equation of 
motion. In deriving the QRPA equations 
two basic approximations  are adopted: (i) The operator,
which determines the excited state, is taken as linear 
superposition of two creation and two annihilation quasiparticle
operators 
by considering the BCS basis as reference. 
(ii) The commutator of bifermion operators is replaced
by its expectation value in the BCS ground state.
This is usually called the "quasiboson approximation" (QBA). 
The QBA violates
the Pauli exclusion principle (PEP) and this affects severely the theory. 
The terms which are left out by the QBA become more and more important 
when the ground state correlations are increased which results in a 
collapse of the
QRPA solution. The approach based on the two approximations
mentioned above will be conventionally called "the standard  QRPA approach". 
 
Recently, the instability of the QRPA solution, caused by the PEP 
violation received much attention from the experts in the field.
In order to improve the reliability of the standard QRPA description
of the nuclear transitions, the renormalized version of the QRPA
(RQRPA), which take into account the PEP in an
approximate way, has been formulated 
\cite{kar93,cat94} and applied to the
$\beta$ and $\beta\beta$ decay problems
\cite{TS95,SSF96,fae98}.  Indeed, the RQRPA does not
collapse within the physical range of the interaction strength
parameters. However, avoiding the collapse 
in the RQRPA a price had to be paid, namely
the violation of the Ikeda sum rule \cite{krm96,hirm96}.

There is a constant interest in studying the physical consequences of 
violating the PEP by  the
QRPA solutions. Some definite conclusions can be drawn by using
solvable models, like, for example, the extensions \cite{hirm96,rad95} to 
proton-neutron systems of
Lipkin or Moszkowski models \cite{lip65,mos58},
as they simulate the realistic cases
either by analytical solutions or by a minimal computational
effort. It is worthwhile mentioning that the study 
of different many-body approximations  within schematic models 
was always of great interest and moreover it is currently considered
of a  major importance 
\cite{hirm96,hirm97,samb97,samb99,hess99,qeq3,dang99,eng97,dob98,pass98,qeq1}. 

The improvement  of the PEP obedience,
within the QRPA, can be achieved in two
ways: (i) By a mapping technique the whole theory can be formulated 
in a boson picture. Such an approach has been outlined 
for the proton-neutron monopole Lipkin model in Ref. 
\cite{samb97,samb99,hess99,qeq3,dang99}.
(ii) One can remain within the fermionic space and derive the
elements of the QRPA equation, at least, perturbatively. Usually
approximations violating the PEP are tested
for Fermi transition since for this case the schematic models 
are simple and solvable 
\cite{hirm96,samb97,samb99,hess99,qeq3,dang99,qeq1}. 
Recently a more general solvable model appeared \cite{eng97,dob98}
that exploits the SO(8) symmetry to include simultaneously the
Fermi and Gamow-Teller transitions. Such a model is devoted
to an extensive treatment  of the proton-neutron pairing interaction.

The goal of this work is to discuss some
limitations of the standard QRPA approach, concerning
the PEP violation.
We shall follow the second possibility mentioned above and
introduce new extensions of the standard QRPA
approach within the proton-neutron monopole Lipkin model 
and point out  some implications for realistic calculations. 
The newly introduced approximations will be compared 
with the exact results, revealing, in this way,
the limits of the approximations. 
Our work completes the discussion of the Refs.
\cite{hirm96,samb97,samb99,hess99,qeq3,dang99} in which
the same schematic model was considered. Indeed, 
for the first time, the QRPA solution with an
exact consideration of the PEP is presented.

The paper is organized as follows. In Sec. II, we describe the 
solvable model and specify the corresponding solution. Section III
describes the standard QRPA and RQRPA within the chosen solvable model. 
In addition, new extensions of the standard QRPA approach,
which take into account the PEP
in an approximate way and exactly are introduced, respectively.
In Sec. IV, the results obtained within the QRPA approaches
are presented and compared with the exact results.
Finally in Sec. V,  we summarize the results and draw some 
conclusions.

%%%%%%%%%%%%%%%%%%%%%%%%%%%%%%%%%%%%%%%%%%%%%%%%%%%%%%%%%%%%%%%%%%%%%%%
%                     Nuclear Hamiltonian                             %
%%%%%%%%%%%%%%%%%%%%%%%%%%%%%%%%%%%%%%%%%%%%%%%%%%%%%%%%%%%%%%%%%%%%%%%

\section{Nuclear Hamiltonian}
\label{sec:level2}

We assume a model  Hamiltonian which includes a single--particle
term,  proton--proton and neutron--neutron  
pairing and a charge-dependent 
two-body interaction with particle--hole and particle--particle channels
included:
\begin{equation} 
H =  H_p + H_n + H_{res},
\label{eq:1}
\end{equation} 
where 
\begin{eqnarray}
H_\tau &=& e_\tau \sum_m a^\dagger_{\tau m}a_{\tau m} 
- G_\tau S^\dagger_\tau S_\tau ~~~(\tau = p, n), ~~~~
\nonumber\\
H_{res} &=& 2~ \chi ~\beta^- ~\beta^+ ~-~ 
2~ \kappa ~ P^- ~P^+,
\label{eq:2}
\end{eqnarray} 
with 
\begin{eqnarray}
S^{\dagger}_{\tau} &=& \frac{1}{2}
\sum_m  a^\dagger_{\tau m}{\tilde a}^\dagger_{\tau m},
\nonumber\\
\beta^- &=& \sum_m a^\dagger_{p m}a_{n m}, ~~~\beta^+ = (\beta^- )^\dagger ,
~~~~~~
\nonumber\\
 P^- &=& \sum_m a^\dagger_{p m}{\tilde a}^\dagger_{n m}, 
~~~P^+ = (P^- )^\dagger ,
\label{eq:3}
\end{eqnarray} 
$a^\dagger$ ($a$) being the particle creation (annihilation) operator and 
$\sim$ indicating the time reversed states
${\tilde a}^\dagger_{\tau m}  = (-1)^{j_\tau - m} 
a^\dagger_{\tau -m}$.

The schematic Hamiltonian, given by Eqs. (\ref{eq:1})--(\ref{eq:3}), 
reproduces well the QRPA results
of the realistic Hamiltonian containing G-matrix elements of the
Bonn-OBEP potential for the beta and double beta decay transitions
\cite{kuz88,civ94,civ95}. The strength $\chi$ ($\kappa$) 
of the particle-hole (particle-particle)
interaction  corresponds to the 
well-known parameter $g_{ph}$ ($g_{pp}$)
commonly used in  literature \cite{vog86,eng88,civ91,mut89}
to parameterize the realistic ph (pp) interaction.

Performing the Bogolyubov transformation
for protons ($\tau = p$) and neutrons ($\tau = n$) 
\begin{eqnarray}
\alpha^\dagger_{\tau m} = u_\tau a^\dagger_{\tau m} - 
v_\tau {\tilde a}_{\tau m}, ~~~~
{\tilde \alpha}_{\tau m} = v_\tau a^\dagger_{\tau m} + 
u_\tau {\tilde a}_{\tau m}, 
\label{eq:4}
\end{eqnarray}
which defines the  quasiparticle representation, and neglecting
the scattering terms $\alpha^\dagger_p \alpha_n$ and 
$\alpha^\dagger_n \alpha_p$, the model Hamiltonian acquires the form
\begin{equation}
H_F = \epsilon C + \lambda_1 A^\dagger A + \lambda_2 
( A^\dagger A^\dagger + A A),
\label{eq:5}
\end{equation}                                                    
with
\begin{eqnarray}
C &=& \sum_m \alpha^\dagger_{p m} \alpha_{p m} + 
\sum_m \alpha^\dagger_{n m} \alpha_{n m}, ~~~~
A^\dagger = [\alpha^\dagger_p \alpha^\dagger_n ]^{J=0},
\nonumber \\
 \lambda_1 &=&  4\Omega [ \chi ( u^2_p v^2_n + v^2_p u^2_n ) - 
                 \kappa (u^2_p u^2_n + v^2_p v^2_n) ]
\nonumber \\
\lambda_2 &=& 4\Omega (\chi + \kappa ) u_p v_p u_n v_n .
\label{eq:6}
\end{eqnarray}
For the sake of simplicity we used a single level case 
$j_p = j_n \equiv j$ and
$G_p = G_n \equiv G$ which implies equal 
energies for protons and neutrons quasiparticles: 
$\epsilon = \epsilon_p = \epsilon_n
= \Omega ~G/2$ and $v_i = \sqrt{N_i/{2\Omega}}$, 
$u_i = \sqrt{1 - N_i/{2\Omega}}$ with $i=p,n$ ($N_p$ and $N_n$ are
 number of protons and neutrons, respectively).
$\Omega$ denotes the semi-degeneracy of the
considered single level. 

The model Hamiltonian in Eqs. (\ref{eq:5}) and (\ref{eq:6}),  
resembles the Hamiltonian
of the Lipkin model \cite{lip65}, when $\lambda_1$ is taken equal to zero.
We note that operators $\{A,A^\dagger  , C\}$ are generators for an 
SU(2) algebra. Indeed their mutual commutators are:
\begin{equation}
[A,A^\dagger] = 1-\frac{C}{2\Omega}, ~~~~~[C,A^\dagger ] = 2 A^\dagger ,
~~~~~~[A,C] = 2 A.
\label{eq:7}
\end{equation}                                                    
This model Hamiltonian is expected to account 
qualitatively for some features of realistic pn-QRPA calculations.
Due to these expectations, it has been 
used to study the standard QRPA, renormalized QRPA as well 
as the higher order QRPA approximations for the many-body system,
Refs.\cite{hirm96,samb97}. The salient feature of this Hamiltonian 
is that the stability of the approximate solutions can be discussed in 
comparison with the exact solution determined by 
diagonalizing $H_F$  in the space of states 
\begin{equation}
|n> = (A^+)^n |0>, ~~~~~~0 \le n \le 2 \Omega. 
\label{eq:8}
\end{equation}
Here $|0>$ denotes the vacuum state for the quasiparticle operators. 
The matrix to be diagonalized can be easily calculated with the
result:
\begin{eqnarray}
<n| H_F |n> &=& 2~ \epsilon ~ n m_n ~+~ 
\lambda_1 ~ ( ~m_{n+1}~ -~ m_n ~+ ~\frac{n~ m_n}{\Omega} ),
\nonumber \\
<n-2| H_F |n> &=& \lambda_2 ~ m_n,
\label{eq:9}
\end{eqnarray}
where
\begin{equation}
m_n \equiv <0| A^n (A^\dagger )^n |0> = 
\frac{ n! ~(2~\Omega)!}{(2~\Omega -n)!~ (2~\Omega)^n } ~~~~~(n \le 2 \Omega ).
\label{eq:10}
\end{equation}
For $n > 2 \Omega$, the norm overlaps  $m_n$ are vanishing.

\section{Quasiparticle Random Phase Approximation }
\label{sec:level3}

Another way to find an excited state for the 
model Hamiltonian (\ref{eq:5}), is to solve the corresponding
QRPA equation. In what follows we shall briefly describe the basic ideas
underlying this method for our solvable model. 

Within the QRPA, an excited state  $|Q>$ is created  by 
applying a phonon creation operator  $Q^\dagger$ on a state 
$|rpa>$ having the properties:
\begin{equation}
|Q> = Q^\dagger |rpa>, ~~~~~~~ Q |rpa> = 0.
\label{eq:11}
\end{equation}                                                    
The simplest form for the phonon operator, in the fermionic space, is
\begin{equation}
Q^\dagger = X A^\dagger - Y A,
\label{eq:12}
\end{equation}                                                    
where X and Y are called forward- and backward- going 
free variational amplitudes and satisfy the  
QRPA equation:
\begin{equation}
\left(
\begin{array}{cc}
{\cal A}&{\cal B}\\
{\cal B}&{\cal A}
\end{array}
\right)
\left(
\begin{array}{c}
X\\
Y
\end{array}
\right)
= {\cal E}_{QRPA}
\left( 
\begin{array}{cc}
{\cal U}&0\\
0&{\cal -U}
\end{array}
\right)
\left(
\begin{array}{c}
X\\
Y
\end{array}
\right),
%\label{dqrpa)
\label{eq:13}
\end{equation}
where
\begin{eqnarray}
{\cal A} &=& <rpa| [ A, [ H_F, A^\dagger ]] |rpa>, \nonumber \\
{\cal B} &=& - <rpa| [ A, [ H_F, A ]] |rpa>, \nonumber \\
{\cal U} &=&  <rpa| [ A, A^\dagger ] |rpa>.
\label{eq:14}
\end{eqnarray}
It is useful to introduce the notation:
\begin{equation}
\overline{X} = {\cal U}^{1/2} X, ~~~~~~\overline{Y} = {\cal U}^{1/2} Y,
\label{eq:15}
\end{equation}
\begin{equation}
\overline{\cal A} = {\cal U}^{-1/2} {\cal A} {\cal U}^{-1/2}, ~~~~
\overline{\cal B} = {\cal U}^{-1/2} {\cal B} {\cal U}^{-1/2}. 
\label{eq:16}
\end{equation}
Then the QRPA eigenenergy ${\cal E}_{QRPA}$ and the new 
amplitudes $\overline{ X}$ and $\overline{ Y}$ 
are given by:
\begin{eqnarray}
{\cal E}_{QRPA} &=& ( {\overline{\cal A}}^2 - {\overline{\cal B}}^2 )^{1/2},
\nonumber \\
\overline{ X} &=& \frac{ \overline{\cal A}+{\cal E}_{QRPA}}
{\sqrt{ {(\overline{\cal A} + {\cal E}_{QRPA})^2} - \overline{\cal B}}^2 },
\nonumber \\
\overline{ Y} &=& \frac{ - \overline{\cal B}}
{\sqrt{ {(\overline{\cal A} + {\cal E}_{QRPA})^2} - \overline{\cal B}}^2 }.
\label{eq:17}
\end{eqnarray}
From the definition of the QRPA ground state $|rpa>$  
(\ref{eq:11}), it follows that elements 
${\cal A}$, ${\cal B}$ and ${\cal U}$ of the QRPA equation 
are function of the $X$ and $Y$ amplitudes. Due to 
this fact this non-linear eigenvalue problem 
could be solved only numerically by an iteration process. The functional
dependence of ${\cal A}$, ${\cal B}$, ${\cal U}$ on 
$X$ and $Y$ is specific to the approximation scheme and influences
crucially the final results. Below, we shall discuss, separately,
several approaches.

\underline{The standard QRPA:} The simplest approximation scheme
to calculate of ${\cal A}$, ${\cal B}$ and ${\cal U}$ is the
quasiboson approximation (QBA), which assumes 
$[A,A^\dagger ] \approx <|[A,A^\dagger ]|>=1$, i.e.
$A$ and $A^\dagger $ are considered to be
boson operators. Here $|>$ denotes the uncorrelated BCS ground state.
In this case, one finds the  expressions:
\begin{equation}
{\cal A} = 2\epsilon + \lambda_1, ~~~~
{\cal B} = 2 \lambda_2, ~~~~
{\cal U} = 1,
\label{eq:18}
\end{equation}
which determine  the excited state  eigenenergy and wavefunction
 with normalization $X^2-Y^2=1$.
The drawback of this approximation scheme is the collapse
of the standard QRPA solution within the physically 
acceptable interval for the nucleon-nucleon interaction
strength. 

\underline{The renormalized QRPA:}
The renormalized QRPA (RQRPA) approach
avoids the collapse of the QRPA solution for  physical
parameters of the nuclear Hamiltonian. Within the RQRPA the
commutator, $[A, A^\dagger ]$ is replaced with its expectation
value in the ground state $D = <rpa|[A,A^\dagger ]|rpa>$
(renormalized QBA). This modifies the matrices
${\cal A}$, ${\cal B}$, ${\cal U}$ 
in the following way \cite{TS95,SSF96}:
\begin{equation}
{\cal A} = 2\epsilon D + \lambda_1 D^2, ~~~~
{\cal B} = 2 \lambda_2 D^2, ~~~~
{\cal U} = D = (1 + \frac{{\overline{Y}}^2}{\Omega})^{-1}. 
\label{eq:19}
\end{equation}
Note that  the fermionic structure of the $A$, $A^\dagger$
operators is taken into account only in an approximate way. In the
limit of $D=1$, i.e. the $|rpa>$ ground state is replaced by the
 BCS one $|>$, one gets the standard QRPA
approach. 

It is worth to remark that in both the standard QRPA and 
the RQRPA, the elements
${\cal A}$, ${\cal B}$ and ${\cal U}$ are evaluated by using some
approximate schemes for the commutator $[A, A^\dagger ]$.
If \underline{the commutator is exactly considered}, i.e. the PEP
 is fulfilled,   the matrices ${\cal A}$, ${\cal B}$ and
${\cal U}$ take the form:
\begin{eqnarray}
{\cal A} &=&  (2\epsilon + \lambda_1 ) - 
(\epsilon + \lambda_1) \frac{<rpa|C|rpa>}{\Omega} + 
\lambda_1 \frac{<rpa|C C|rpa>}{4 \Omega^2} - \nonumber \\
&& \lambda_1 \frac{<rpa| A^\dagger A |rpa>}{\Omega} - 
2 \lambda_2 \frac{<rpa| A^\dagger A^\dagger|rpa>}{\Omega}, 
\label{eq:20}
\end{eqnarray}
\begin{eqnarray}
{\cal B} &=&  \lambda_2 (2 - \frac{1}{\Omega} ) -
\lambda_2 (2 - \frac{1}{2 \Omega} ) \frac{<rpa|C|rpa>}{\Omega} + 
\lambda_2 \frac{<rpa|C C|rpa>}{2 \Omega^2} - \nonumber \\
&&2 \lambda_2 \frac{<rpa| A^\dagger A |rpa>}{\Omega} - 
\lambda_1 \frac{<rpa| A A |rpa>}{\Omega},
\label{eq:21} \\
{\cal U} &=&  1 - \frac{<rpa|C|rpa>}{2 \Omega}. 
\label{eq:22} 
\end{eqnarray}
The calculation of the involved matrix elements
requires the knowledge of the $|rpa>$ ground state,
determined by the condition in Eq. (\ref{eq:11}). The analytical
form for $|rpa>$  is known 
within  QBA and renormalized QBA.
For the phonon operator given by the Eq. (\ref{eq:12}), one obtains:
\begin{equation}
|rpa>_{_{QBA}} = n e^{-d A^\dagger A^\dagger} |>,~~~~~
d = - \frac{Y}{2 X},
\label{eq:23}
\end{equation}
where $n$ stands for the normalization factor.

In general, it is not possible to find an explicit expression
for $|rpa>$ unless some additional approximation is adopted.
Fortunately, this can be achieved in the case of the solvable model
considered in the present paper.
By solving the Eq. (\ref{eq:11}), one 
obtains\footnote{
After the present paper was completed we learned that this equation has been 
derived also by other authors \cite{qeq0,qeq1,qeq3}} 
\begin{equation}
|rpa>_{_{exc.}} = N \sum_0^{\Omega} \beta_l (\frac{Y}{X})^l 
(A^\dagger A^\dagger )^l |>, 
\end{equation}
with
\begin{eqnarray}
\beta_l = (2\Omega)^l \frac{\Omega !}{(2 \Omega )!}
\frac{(2 \Omega -2 l)!}{l! (\Omega - l)! }, 
~~~~~N^{-2} = \sum_0^\Omega \beta^2_l (\frac{Y}{X})^{2 l}
m_{2 l}.
\label{eq:24}
\end{eqnarray}

We note that if the ground state correlations  are neglected 
in the calculation of
${\cal A}$, ${\cal B}$ and ${\cal U}$ in 
Eqs. (\ref{eq:20})--(\ref{eq:22}), i.e. the RPA vacuum 
$|rpa>$ is replaced with BCS vacuum $|>$, one gets:
${\cal A} =  (2\epsilon + \lambda_1 )$, 
${\cal B} =  \lambda_2 (2 - \frac{1}{\Omega} )$ and
${\cal U} =  1$.  Obviously, the standard QRPA equation is recovered
in the limit $\Omega \rightarrow \infty$, i.e. 
the bifermion operators $A^+$ and $A$ behave like bosons.

By using the approximate $|rpa>_{_{QBA}}$ (\ref{eq:23})
and exact $|rpa>_{_{exc.}}$ (\ref{eq:24})
solutions for the QRPA ground
state, one achieves, in fact, 
new extensions of the standard QRPA approach, namely
the \underline{QRPA with the PEP (PP QRPA) included in an approximate manner}
and the \underline{QRPA with the PEP (EPP QRPA) fully fulfilled},
respectively. 
Both  methods go beyond the renormalized
QRPA approach and require to evaluate ground state
expectations values of the $C$, $CC$, $A^\dagger A^\dagger$ and 
$A^\dagger A$ operators entering the expression for ${\cal A}$, 
${\cal B}$ and ${\cal U}$ as shown in Eqs. (\ref{eq:20})--(\ref{eq:22}).
This is performed free of any approximation.

\underline{The QRPA with PEP:} 
There is an important difference
between this method and the RQRPA one, although both methods use 
the same ground state wave functions [see Eq. (\ref{eq:22})]. 
Indeed, within the RQRPA, the commutator
of the two bifermion operators $A$ and $A^\dagger$ is considered
in approximate way while the PP QRPA takes it exactly. Indeed,
the operators of interest have the expectation values:
\begin{eqnarray}
_{_{QBA}}<rpa| C |rpa>_{_{QBA}} & = & -4 ~n^2~d ~h_2(d), 
\nonumber \\
_{_{QBA}}<rpa| C C |rpa>_{_{QBA}} &=& - 16~n^2  
(d~ h_2(d) ~- ~ d^2~ h_4(d)), 
\nonumber \\
_{_{QBA}}<rpa| A^\dagger A^\dagger |rpa>_{_{QBA}} &=& 
 ~n^2~h_2(d), \nonumber \\
_{_{QBA}}<rpa| A^\dagger A |rpa>_{_{QBA}} &=& 
-~n^2( (2 -\frac{1}{\Omega}) ~d~ h_2(d) 
+ 2 ~\frac{d^2}{\Omega} ~h_4(d)).
\label{eq:25}
\end{eqnarray}
where the following notations have been used:
\begin{eqnarray}
h_0 (d) & \equiv &  <| e^{-d A A} e^{-d A^\dagger A^\dagger} |>
 = \frac{1}{n^2}, \nonumber \\
&& =\sum_{j=0}^\Omega ~\frac{d^{2 j}}{(j!)^2}~ m_{2 j} 
\approx m_0 + d^2 m_2  + \frac{d^4}{4} m_4,
\label{eq:26}
\end{eqnarray}
\begin{eqnarray}
h_2 (d) & \equiv &  <| e^{-d A A}A^\dagger 
A^\dagger e^{-d A^\dagger A^\dagger} |>, \nonumber \\
&& =\sum_{j=0}^{\Omega -1} ~\frac{d^{2 j}}{(j!)^2} ~\frac{-d}{j+1}~ m_{2 j+2}
 \approx -d m_2 - \frac{d^3}{2} m_4 ,
\label{eq:27}
\end{eqnarray}
\begin{eqnarray}
h_4 (d) & \equiv &  <| e^{-d A A} (A^\dagger )^4 
e^{-d A^\dagger A^\dagger} |>, \nonumber \\
&& =\sum_{j=0}^{\Omega -2} ~\frac{d^{2 j}}{(j!)^2} ~\frac{d^2}{(j+2)(j+1)}~ 
m_{2 j+4} \approx \frac{d^2}{2} m_4,
\label{eq:28}
\end{eqnarray}
We hope  that this approach can be applied also
for realistic calculations and within a large model space. 
Note that knowing $|rpa>_{_{QBA}}$, 
the QRPA matrices can be evaluated  without  the PEP violation,
at least perturbatively with respect to the factor $d$. 
If the pertubative series is truncated to the quadratic terms in d,
the resulting approach will be hereafter labeled by the abbreviation PP2 QRPA.

\underline{The QRPA with exact PEP:} 
This method can be formulated
only for a solvable model for which the exact QRPA ground state
can be analytically found. Within the EPP QRPA there is no violation
of the PEP. However, one can not expect that the EPP QRPA solution
coincides with the exact solution for the
first excited state of the nuclear Hamiltonian  
(\ref{eq:5}). The difference is caused by 
the approximations incorporated
in the construction of the operator $Q^\dagger$ determining
the excited state [see Eq. (\ref{eq:11})]. Therefore, from the direct
comparison with the exact solution one may conclude how far the 
approximate description, with the phonon operator of a simple structure, is
from the exact picture (\ref{eq:11}). The following expressions are used
in elaborating the above defined procedure:
\begin{eqnarray}
_{_{exc.}}<rpa|A A|rpa>_{_{exc.}} &=& 
_{_{exc.}}<rpa|A^\dagger A^\dagger |rpa>_{_{exc.}} = 
N^2 \sum_{l=0}^{\Omega -1} \beta_l \beta_{l+1}
(\frac{Y}{X})^{2l+1} m_{2l+2}, \nonumber \\
_{_{exc.}}<rpa|A A^\dagger |rpa>_{_{exc.}} &=&  
N^2 \sum_{l=0}^{\Omega -1} \beta^2_l 
(\frac{Y}{X})^{2l} m_{2l+1}, \nonumber \\
_{_{exc.}}<rpa| C |rpa>_{_{exc.}} &=&  
N^2 \sum_{l=1}^{\Omega} \beta^2_l (4l) 
(\frac{Y}{X})^{2l} m_{2l}, \nonumber \\
_{_{exc.}}<rpa| C C |rpa>_{_{exc.}} &=&  
N^2 \sum_{l=1}^{\Omega} \beta^2_l (4l)^2 
(\frac{Y}{X})^{2l} m_{2l}, \nonumber \\
_{_{exc.}}<rpa|A^\dagger A |rpa>_{_{exc.}} &=& -1 + 
\frac{_{_{exc.}}<rpa| C |rpa>_{_{exc.}}}{2\Omega}
+ _{_{exc.}}<rpa|A A^\dagger |rpa>_{_{exc.}}.
\label{eq:29}
\end{eqnarray}

\section{RESULTS AND DISCUSSIONS}
\label{sec:level4}

In what  follow we shall present the numerical results for the 
QRPA approaches described in the previous section and 
compare them  with the values provided by diagonalizing
$H_F$ (\ref{eq:5}). In order to continue and complete
the discussion of this Hamiltonian given 
in Refs. \cite{hirm96,samb97}, we have chosen the same set of 
parameters  as there:
\begin{equation}
j = 9/2, ~~~~~ Z = 4, ~~~~~~ N = 6, ~~~~~e = 1 ~MeV,
\end{equation}
which determine the BCS amplitudes entering the 
$lambda_{1,2}$ parameters of the model Hamiltonian $H_F$.
Also we redefine the parameters $\kappa$ and $\chi$ as 
in Refs. \cite{hirm96,samb97}:
\begin{equation}
\kappa \rightarrow {\kappa}' \equiv 2\Omega~ \kappa, ~~~~~
\chi \rightarrow {\chi}' \equiv 2\Omega ~ \chi.
\end{equation}

The values ${\chi}'=0.$ 
and ${\chi}'=0.5$  were adopted while the particle-particle
strength ${\kappa}'$ was allowed to vary in the interval 
$0 \leq {\kappa}' \leq 2$. Comparing the
schematic calculations with the realistic ones, 
a value for  ${\kappa}'$ close to unity is expected. 

\subsection{Excitation energies}

In Fig. \ref{fig.1} we plotted the dependence of the QRPA excitation
energy and the first excitation energy obtained by
diagonalizing   $H_F$ (bold solid line) 
on ${\kappa}'$, for ${\chi}'=0$ (upper figure) and 
${\chi}'=0.5$ (lower figure), respectively. Note that  the standard QRPA 
breaks down for 
${\kappa}'\approx 1.$ The RQRPA excitation energy remains real
within the whole interval of ${\kappa}'$, although it
deviates significantly from the exact solution 
beyond the breaking down point of the standard QRPA.
The effect is more evident for ${\chi}'=0.5$. 
The PP2 QRPA and, especially, the PP QRPA
energies reproduce quite well those of $H_F$, except for the values
of ${\kappa}'$ approaching their minimum. 
The EPP QRPA, which take into account the PEP
exactly, systematically overestimates the results obtained through
the diagonalization of $H_F$. This difference might be attributed
to the simple form of the phonon operator $Q^\dagger$ 
(\ref{eq:12}). From this figure one remarks that the collapse is 
shifted to a large value of ${\kappa}'$ when the PEP is 
satisfied to a larger extent. For example for the PP2 QRPA, PP QRPA and 
EPP QRPA, the collapse appear at about  ${\kappa}'=1.80$, 
${\kappa}'=2.50$ and  ${\kappa}'=2.55$  
(${\chi}'=0.5$), respectively. This indicates that a real phase
transition could take place in the region beyond ${\kappa}'=2.5$.

\subsection{Expectation values of the quasiparticle number operator}

In order to get additional information about the quality of different
approximations, we calculate the expectation values of the
quasiparticle number operator  in the ground and first 
excited states. These are defined as follows:
\begin{equation}
N_0 \equiv <rpa| \frac{C}{2} |rpa>, ~~~~
\Delta N  \equiv  <rpa| Q \frac{C}{2} Q^\dagger |rpa> - N_0.
\end{equation}
For the situations defined before, the results are:
\begin{eqnarray}
N_0 &=& Y^2,~~~~~~~~~~~~~~~~~~~~~~(QRPA),\nonumber \\
&=& {\overline{Y}}^2,~~~~~~~~~~~~~~~~~~~~~~(RQRPA),\nonumber \\
&=& \frac{2 d^2 m_2}{1+d^2 m_2}, ~~~~~~~~(PP2~ QRPA)
\nonumber \\
&=& n^2 ~(-2 d)~h_2(d),~~~~~~~~~~~~~~~~(PP~ QRPA), \nonumber \\
&=& _{_{exc.}}<rpa| C |rpa>_{_{exc.}}, ~~~~~~~~~(EPP~ QRPA)
\end{eqnarray}

\begin{eqnarray}
\Delta N &=& 1+2Y^2,~~~~~~~~~~~~~~~~~~~~~~~~~~~~(QRPA), \nonumber \\
& =& 1+2{\overline{Y}}^2,~~~~~~~~~~~~~~~~~~~~~~~(RQRPA),\nonumber \\
 &=& (X^2+Y^2) + (X^2 - Y^2 - 1 -\frac{X^2+Y^2}{\Omega})
\frac{_{_{QBA}}<rpa| C |rpa>_{_{QBA}}}{2}- \nonumber \\
&& {\frac{X^2-Y^2}{4 \Omega}}  
{_{_{QBA}}<rpa| C C |rpa>_{_{QBA}}} ~~~~~~~(PP~ QRPA), \nonumber \\
 &=& (X^2+Y^2) + (X^2 - Y^2 - 1 -\frac{X^2+Y^2}{\Omega})
\frac{_{_{exc.}}<rpa| C |rpa>_{_{exc.}}}{2}- \nonumber \\
&& {\frac{X^2-Y^2}{4 \Omega}}  
{_{_{exc.}}<rpa| C C |rpa>_{_{exc.}}} ~~~~~~~(EPP~ QRPA).
\end{eqnarray}
The expressions corresponding to 
the standard QRPA and RQRPA have been obtained
by replacing the operator $C/2$ with its boson image, respectively.
As for the remaining case we stay in
the fermionic space and use the commutation algebra given by
the Eq. (\ref{eq:7}). 

In Figs. \ref{fig.2} and \ref{fig.3}, $N_0$ and $\Delta N$, given by
the above listed approximation schemes as well as the exact 
calculation are presented.
One sees that the standard QRPA overestimates the ground state
correlations near the collapse point, which reflects itself in a
sudden increase of the average quasiparticle number. 
 The RQRPA does not collapse at all and
overestimates the exact result both for $N_0$ and $\Delta N$.
A distinct situation is produced by the QRPA approaches with PEP.
There occurs an underestimation of the ground state correlations
and the values of $N_0$ are smaller than those corresponding to the
exact solution. Practically, there is no difference between the
PP2 QRPA and the PP QRPA up to the point where the former one collapses. 
The EPP QRPA, with an exact treatment of the PEP, provides
the best agreement with the exact results  obtained by  diagonalizing
$H_F$. We notice that none of the considered QRPA methods is able
to reproduce the exact result for $\Delta N$ which, after a certain
point in the region ${\kappa'} \doteq 1.-1.1$, is falling down,
contrary to the behavior of other curves\cite{samb97}. This is an indication that 
a more complex form for the phonon operator might be necessary.

\subsection{Fermi $\beta$ transition amplitudes} 

We turn our attention now to transitions induced by the 
Fermi $\beta^\pm$ operators. Neglecting the scattering term, 
as we did for the nuclear Hamiltonian $H_F$
(see Sec. II), the Fermi $\beta^\pm$ operators, in
the quasiparticle basis, take the form
\begin{equation} 
\beta^-  = \sqrt{2\Omega} ( u_p v_n A^\dagger + v_p u_n A ),
~~~~~\beta^+ = (\beta^- )^\dagger .
\end{equation}
The matrix elements of $\beta^-$ operator between ground
and first excited states corresponding to  different versions
of the QRPA, are given as follows: 
\begin{eqnarray}
<0^+_1|\beta^- |rpa> &=& \sqrt{2\Omega} (X u_p v_n + Y v_p u_n )
~~~~~~~ (QRPA),\nonumber \\
&=& \sqrt{2\Omega} (\overline{X} u_p v_n + \overline{Y} v_p u_n )
D^{1/2}~~~~~~~ (RQRPA),\nonumber \\
&=&  \sqrt{2\Omega} (X u_p v_n + Y v_p u_n )
\frac{1 + d^2 (1-\frac{2}{\Omega}) m_2}{1+d^2 m_2}
~~(PP2~QRPA),\nonumber \\
&=& \sqrt{2\Omega} n^2 (X u_p v_n + Y v_p u_n )
( h_0(d) + 2 \frac{d}{\Omega} h_2(d) )  ~~~ (PP~QRPA), \nonumber \\
&=& \sqrt{2\Omega} (X u_p v_n + Y v_p u_n ) \times \nonumber \\ 
&&~~~~~ (1- \frac{_{_{exc.}}<rpa| C |rpa>_{_{exc.}}}{2 \Omega}) ~~~
(EPP~ QRPA).
\end{eqnarray}
Expressions for $\beta^+$ are obtainable from the above equations by
interchanging the  $u'$s and $v'$s.

In Figs. \ref{fig.4} and \ref{fig.5} we examine 
the behavior of the
$\beta^-$ and $\beta^+$ amplitudes respectively
as functions of ${\kappa}'$
for two values for ${\chi}'$ ($= 0., ~0.5$). One notes a rapid
increase of  $\beta^-$ and decrease  of $\beta^+$ transition strengths
for the standard QRPA and the PP2 QRPA, close to their
 collapse  point. 
In general, the PP2 QRPA, the PP QRPA and the EPP QRPA methods 
reproduce better the trends of the exact results 
comparing them with the RQRPA. The best agreement 
with the exact values is achieved  for the PP QRPA. Obviously these 
analysis demonstrate how important it is to have a correct treatment
of the PEP, in describing the nuclear $\beta$ transitions.
It is worth noting the sensitivity of the $\beta^+$ 
amplitude to the details of nuclear structure wave functions. 
Indeed, by increasing the
the particle-particle interaction strength
${\kappa}$, the matrix element of the $\beta^+$ transition operator
reaches a vanishing value and therefore is no longer predictable.  

\subsection{The Ikeda sum rule}

From the $\beta^\pm$ amplitudes one obtains, straightforwardly,
the $\beta^\pm$ strengths:
\begin{equation}
S^- = |<0^+_1|\beta^- |rpa>|^2, ~~~~
S^+ = |<0^+_1|\beta^+ |rpa>|^2.
\end{equation}
For a ground state preserving the proton and neutron numbers 
in average, 
the two strengths are satisfying the Ikeda sum rule
\begin{equation}
S^- - S^+ = N - Z
\end{equation}
where N and Z are the numbers of neutrons and protons, respectively.

It is well known that the Ikeda Sum rule is automatically
fullfiled in the standard QRPA and violated in the RQRPA
\cite{krm96,hirm96}. According to the 
Fig. \ref{fig.6}, this is also true
for the solvable  model of this work.
This figure shows also numerical results for the
Ikeda sum rule predicted by the 
methods described so far,
and compare them with the exact results. One notices that the
exact results for $H_F$, do not fulfill the Ikeda sum rule 
and show a large deviation from the value of (N-Z) for
$\kappa'$ for $\kappa' \ge 1$. 
The origin of this phenomenon is expected
to due to neglecting the scattering terms in the derivation
of the assumed Hamiltonian $H_F$ \cite{rad98}. 
We hope that discrepancies concerning the Ikeda sum rule
are substantially diminished by adding the contributions due 
to the quasiparticle operators $a^\dagger_\tau a_\tau$,
$a^\dagger_\tau a^\dagger_\tau + a_\tau a_\tau$ involved
in the particle number operators. Indeed these contributions have 
been omitted so far, although their average on the ground state is
not vanishing.
One may conclude that the standard QRPA
fails to reproduce the exact results for the chosen Hamiltonian. 
All other  modifications of the QRPA (RQRPA,
PP2 QRPA, PP QRPA, EPP QRPA) reproduce the trend of the exact
solution, albeit the agreement with the exact results, 
for large value of $\kappa'$
($\kappa' \ge 1$) is rather poor. For non-vanishing $\kappa'$ the results
lying closest to the exact ones are those produced by PP QRPA and FPP QRPA.
A reason might be the influence of higher excited states on the
ground state induced by the diagonalization procedure. It is 
worthwhile to notice that improving the treatment of the 
PEP in the QRPA, a better agreement of the Ikeda sum rule 
with the (N-Z) value is achieved.

\subsection{Double beta decay matrix element}

In this section we shall focus our attention on the two-neutrino
double beta decay mode, $2\nu\beta\beta$. 
Consequences of the
previously presented approaches on the $2\nu\beta\beta$-decay
matrix element will be discussed.
Within the solvable model considered here
there is only one QRPA excited state and the 
corresponding nuclear $2\nu\beta\beta$-decay matrix element
takes the form
\begin{equation}
M_F^{2\nu} = \frac{{_f}<rpa|\beta^- |0^+_1>_f  
{_i}<0^+_1|\beta^- |rpa>_i}{ {\cal E}_{QRPA} +\Delta}.
\end{equation}
Here, the states with subscripts "i" and "f" are describing 
to the initial (A,Z) and  final (A,Z+2)  nuclei, respectively.
We considered $\Delta$ to be equal to 0.5 MeV and performed
the calculations for the following set of parameters \cite{hirm96}:
\begin{equation}
j=19/2~ ~(~Z=6,~~N=14)~~\rightarrow ~ ~
(~Z+2 = 8,~~N-2 = 12),~~~  \chi'=0,~0.5
\end{equation}

Results corresponding to the matrix element $M^{2\nu}_F$,
calculated with different approximations,
 are shown in Fig. \ref{fig.7}, as  function of 
$\kappa'$. In addition we present also the exact results of $H_F$,
considering only the contribution coming from the lowest 
intermediate state. One  notices that the behaviors of 
the QRPA and the RQRPA curves are qualitatively similar 
to those found in the
realistic calculations. The transition amplitude
 $M^{2\nu}_F$ vanishes
within the range $1.0 \le \kappa' \le 1.5$ for all calculations,
including the exact ones. 
The values of $M^{2\nu}_F$  obtained by a better consideration
of the Pauli exclusion principle  (PP QRPA and EPP QRPA) are
significantly less sensitive to the
particle-particle interaction strength $\kappa'$. Our studies show that 
the behavior of $M^{2\nu}_F$, as function of $\kappa'$, is strongly
influenced by the $\beta^+$ strength characterizing final nucleus.

\section{SUMMARY AND CONCLUSIONS}
\label{sec:level6}

We have analyzed some limitations of the QRPA formalism
in an solvable 
proton-neutron monopole Lipkin model. In addition to the
standard QRPA and RQRPA, we introduced new extensions of the 
standard QRPA approach by improving (PP QRPA) and exactly 
(EPP QRPA) considering  the Pauli exclusion 
principle\footnote{
After the present work was completed  
we were aware  about a certain overlap of the method -  EPP 
QRPA presented here and the self-consistent QRPA approach 
proposed in Ref. \cite{qeq1}. However there are also 
important differences such as i) the u and v coefficients are 
determined differently ii) the model Hamiltonians are different 
iii) our papers have distinct objectives.} 
These approaches have been used to study 
the behavior of different observables as
function of the particle-particle interaction strength $\kappa'$.

Due to the collapse of its
solution for $\kappa' \approx 1.0$, 
the standard QRPA reproduces worse the exact
results of the nuclear Hamiltonian. Our studies show 
that the real collapse of the QRPA, usually associated 
to a phase transition,
appears for a large value of $\kappa'$ ($\approx 2.5$). For the 
PP2 QRPA, which is obtained from the PP QRPA by cutting the series
in d at second  order, the solution breakdown  appears
around $\kappa' =1.7$. This suggests that a better consideration
of the PEP, within the QRPA, shifts the instability strength
to a larger value.

To shed more light on this problem we commented on the 
corrections induced by the ground state correlations, by
plotting the average quasiparticle  number versus $\kappa'$.
Our analysis of the $\beta^\pm$ transitions shows 
the sensitivity of the $\beta^+$ transition to correlations,
included in the ground state, which violate the PEP.
Concerning the
Ikeda sum rule we have found that this is  conserved neither
by the exact solution of the quasiparticle nuclear Hamiltonian 
nor by the QRPA methods including the PEP. 
This discrepancy  is very likely due to the structure of the model
Hamiltonian.
Also one expects, that including the contribution coming from the two 
quasiparticle and quasiparticle scattering operators entering the expressions
of the particle number operators, the discrepancies are decreased.

The $2\nu\beta\beta$-decay matrix element, calculated 
within the solvable model, is changing the sign when
$\kappa'$ is increased, similar to what happens in realistic calculations. 
It was pointed out that $M^{2\nu}_F$ 
 calculated by the PP QRPA and EPP QRPA, 
is less sensitive to the details of nuclear structure
than that predicted by 
the standard QRPA and RQRPA approaches.

In discussing the results of the PP2 QRPA, PP QRPA and EPP 
QRPA we should keep in mind that the first two methods are just approximations
of the last method. As expected, the PP QRPA  reproduces 
the EPP QRPA solution better than the PP2 QRPA since it uses a complete 
pertubative expansion. The difference between the EPP QRPA and
exact solution originates in the simple form of the
phonon operator which ignores nonlinear terms like for example those
proportional to the $A^+A^+A$ and $A^+AA$ operators. To our knowledge, 
we are the first showing the inaccuracy 
coming exclusively from a simple form of the widely used 
phonon operator in the QRPA. An interesting point is the connection
of the EPP QRPA and the PP QRPA to the exact solution. We note that
there is no principle reason which would determine that the EPP QRPA 
results should 
reproduce the exact ones better than the PP QRPA. 
It turns out that the combined effect coming from the simple form 
of the phonon operator 
and the simple RPA wave functions, derived within 
quasiboson approximations, leads to a better agreement with exact solution
for PP QRPA 
in some cases, e.g. for the lowest excitation energy 
(Fig. \ref{fig.1}) and $\beta^+$ transition amplitudes (Fig. \ref{fig.5})
and worse for other observables. 
For example the EPP QRPA reproduces the exact results better than PP QRPA
when the expectation 
values of the quasiparticle number operators (Figs. \ref{fig.2} and
\ref{fig.3}) and Ikeda Sum Rule (Fig. \ref{fig.6}) are calculated.

The main conclusions of our analysis can be summarized as follows. 

New extensions of the standard QRPA with approximate 
(PP QRPA) and exact (EPP QRPA)
consideration of the PEP were presented. These formalisms yield a
better agreement 
with the exact results, obtained by diagonalizing 
the model Hamiltonian, than the standard QRPA and the renormalized 
QRPA approaches. 

The EPP QRPA
results show that the collapse of the first excited states 
is far from the place where the standard QRPA breaks down,
i.e., is achieved for 
larger values of the particle-particle
interaction strength. 

The comparison of the EPP QRPA results
with those obtained with the exact eigenstates of $H_F$, 
points out
the drawbacks coming from the simple
structure of the QRPA phonon operator
and suggests a range of applicability for this theory.
Clearly, this analysis shows some limitations for the QRPA 
and RQRPA approaches.

The results of the present paper support our hope that 
the PP QRPA approximations might work equally well  in the
case of realistic calculations for large model space,
which are expected to offer more reliable results for the
double beta decay transitions than the standard
QRPA and the RQRPA ones. Indeed, the PP QRPA
is based on the  approximate QRPA ground state wave functions, derived 
within the QBA, which  can be undoubtly found
also in realistic models.
This subject seems to be very interesting and therefore deserves
further considerations. We intend to apply the PP2 QRPA method first
for a realistic models with separable forces \cite{kuz88}.

\acknowledgments
We acknowledge a partial support from the CNSIS
under the contract 39C (A. A. R.), 
the Deutsche Forschungsgemeinschaft 
Fa67/17-1 and Fa67/19-1 and the Grant Agency of the Czech 
Republic grant No. 202/98/1216 (F. \v S).

%%%%%%%%%%%%%%%%%%%%%%%%%%%%%%%%%%%%%%%%%%%%%%%%%%%%%%%%%%%%%%%%%%%%%%%%
%%%%%%%%%%%%%%%%%          Figures section          %%%%%%%%%%%%%%%%%%%% 
%%%%%%%%%%%%%%%%%%%%%%%%%%%%%%%%%%%%%%%%%%%%%%%%%%%%%%%%%%%%%%%%%%%%%%%%

\begin{figure}
%\vspace{-1.8cm}
%\centerline{\epsfig{file=nfig1.ps,height=20.cm}}
%\vspace{-1.5cm}
\caption{Comparison of the lowest excitation energies resulting
from the diagonalization of $H_F$ (bold solid line - $H_F$ diag.) 
with the standard QRPA (solid line - QRPA), the renormalized QRPA 
(dashed line - RQRPA), the QRPA with the Pauli exlusion 
principle up to second order in d (bold dashed line - PP2 QRPA), 
the QRPA with Pauli exclusion principle (bold dotted line - PP QRPA) and the 
QRPA with exact consideration of Pauli exclusion principle 
(bold dot-dashed line - EPP QRPA)
values. 
}
\label{fig.1}
\end{figure}

\begin{figure}
%\vspace{-1.8cm}
%\centerline{\epsfig{file=nfig2.ps,height=20.cm}}
%\vspace{-1.5cm}
\caption{The expectation values of (half) the
quasiparticle total number operator, C/2, in the ground state,
versus $\kappa'$. Conventions are the same as in Fig.1.
}
\label{fig.2}
\end{figure}

\begin{figure}
%\vspace{-1.8cm}
%\centerline{\epsfig{file=nfig3.ps,height=22.cm}}
%\vspace{-1.5cm}
\caption{Differences between the expectation values of 
(half) quasiparticle number operator, C/2, in the first excited 
state and in the ground state as function of $\kappa '$. 
Conventions are the same as in Fig.1.
}
\label{fig.3}
\end{figure}

\begin{figure}
%\vspace{-1.8cm}
%\centerline{\epsfig{file=nfig4.ps,height=22.cm}}
%\vspace{-1.5cm}
\caption{The Fermi $\beta^-$ transition amplitudes between the 
ground and first excited states. 
Conventions are the same as in Fig.1.
}
\label{fig.4}
\end{figure}

\begin{figure}
%\vspace{-1.8cm}
%\centerline{\epsfig{file=nfig5.ps,height=22.cm}}
%\vspace{-1.5cm}
\caption{The Fermi $\beta^+$ transition amplitudes between the 
ground and first excited states. 
Conventions are the same as in Fig.1.
}
\label{fig.5}
\end{figure}

\begin{figure}
%\vspace{-1.8cm}
%\centerline{\epsfig{file=nfig6.ps,height=22.cm}}
%\vspace{-1.5cm}
\caption{The Fermi $\beta^-$ transition amplitudes between the 
ground and first excited states. 
Conventions are the same as in Fig.1.
}
\label{fig.6}
\end{figure}

\begin{figure}
%\vspace{-1.8cm}
%\centerline{\epsfig{file=nfig7.ps,height=22.cm}}
%\vspace{-1.5cm}
\caption{The $2\nu\beta\beta$-decay Fermi transition amplitude
versus $\kappa '$. The same notations as in Fig. 1 are used.
}
\label{fig.7}
\end{figure}
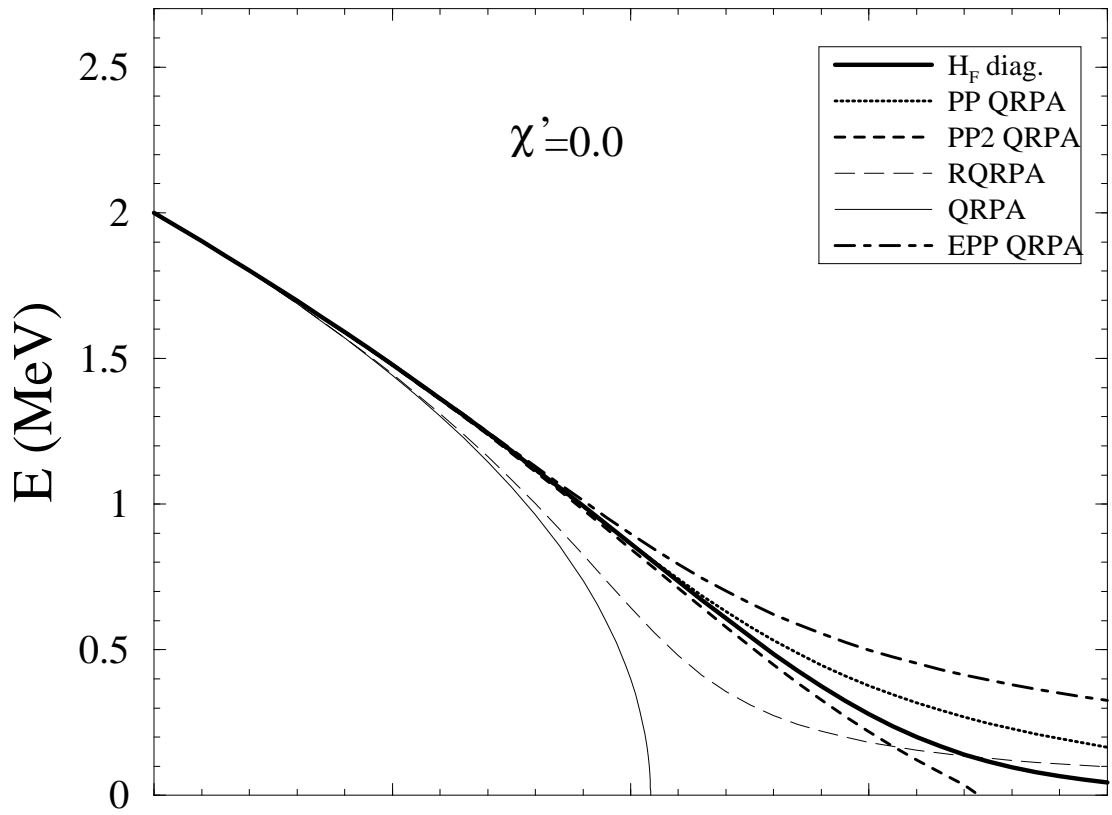
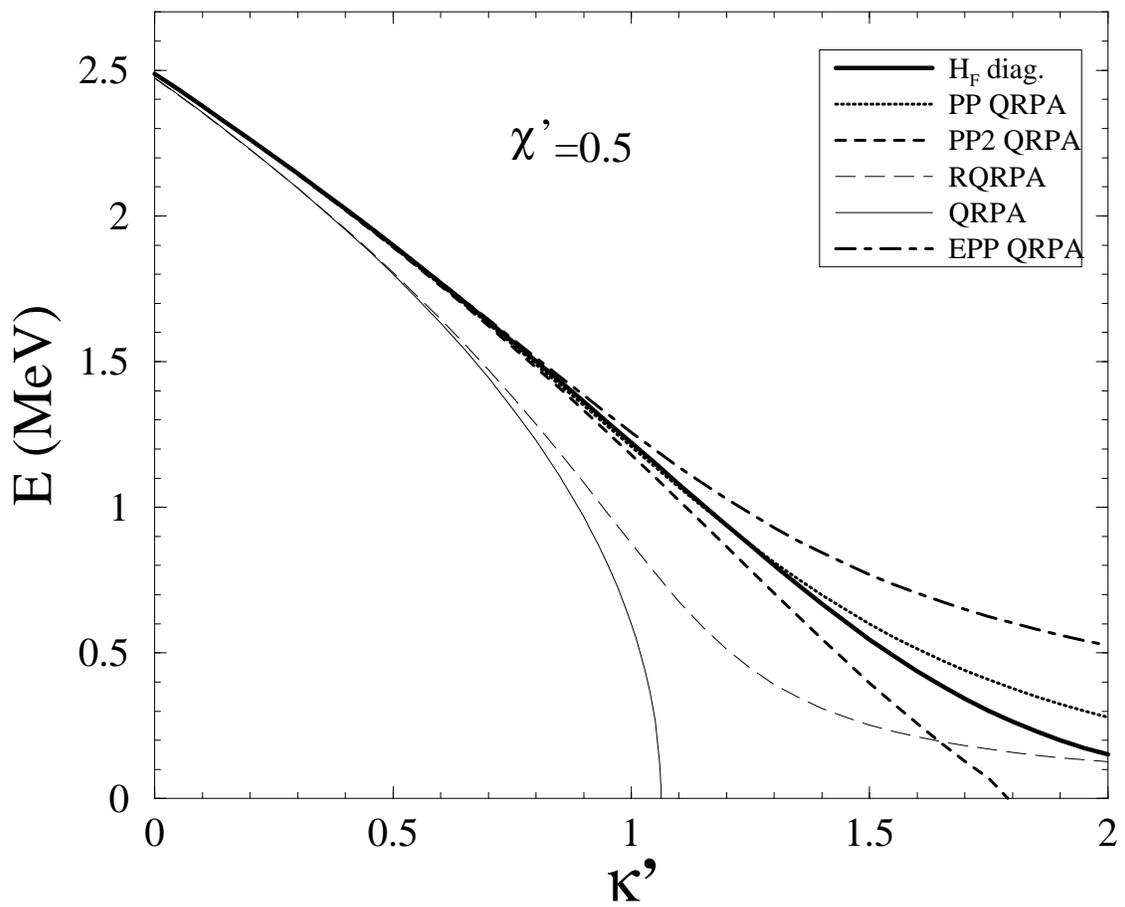
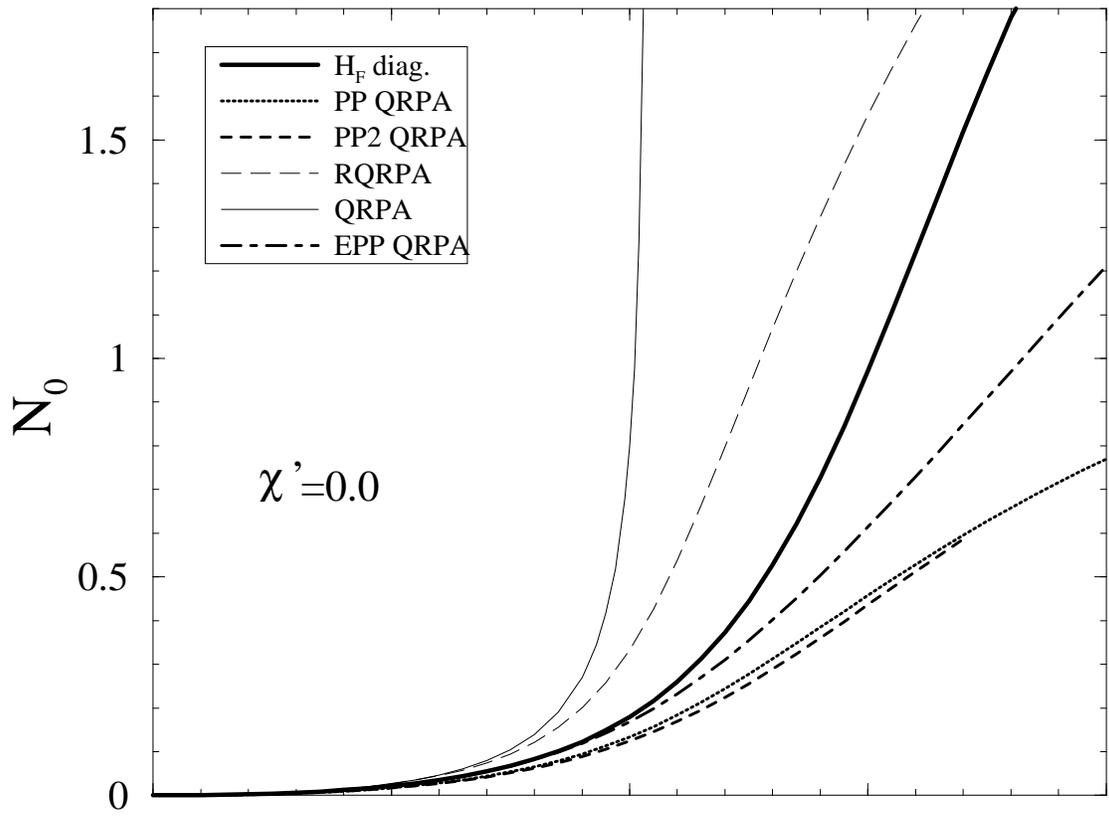
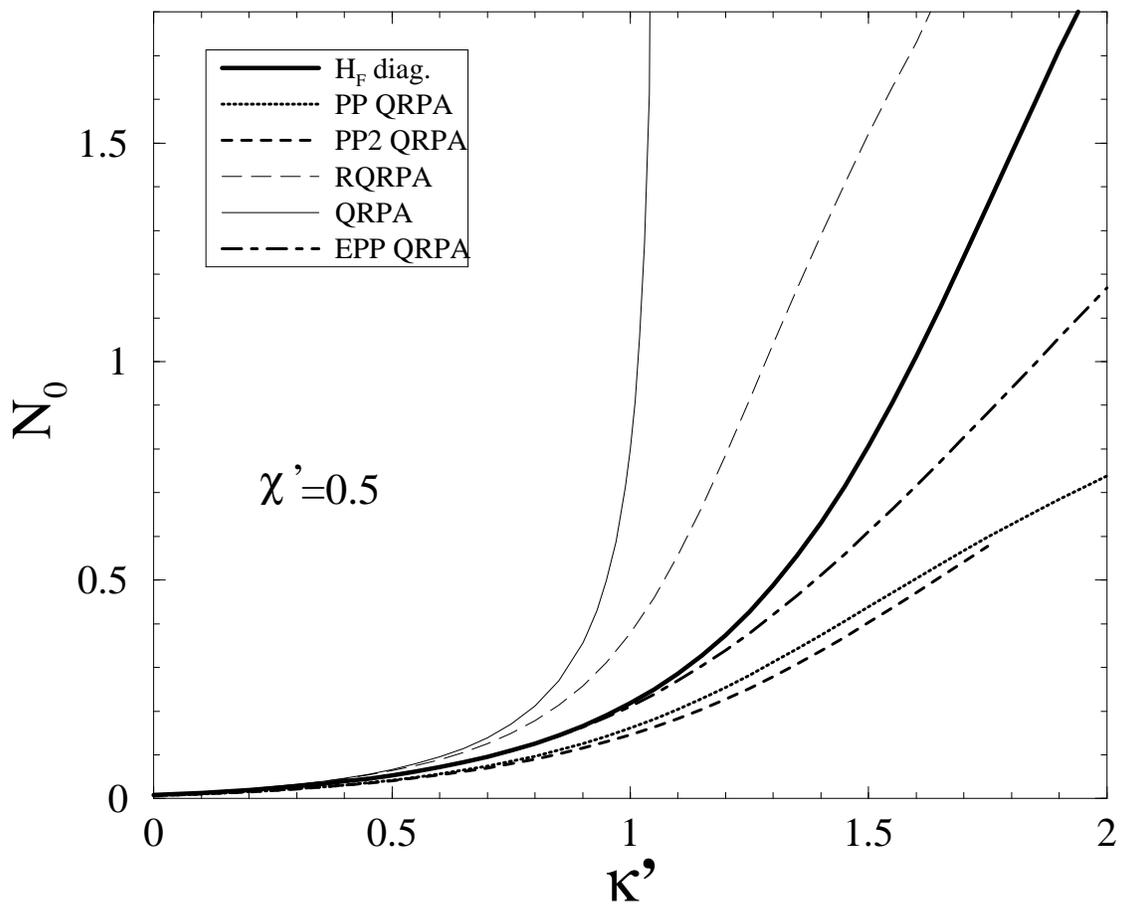
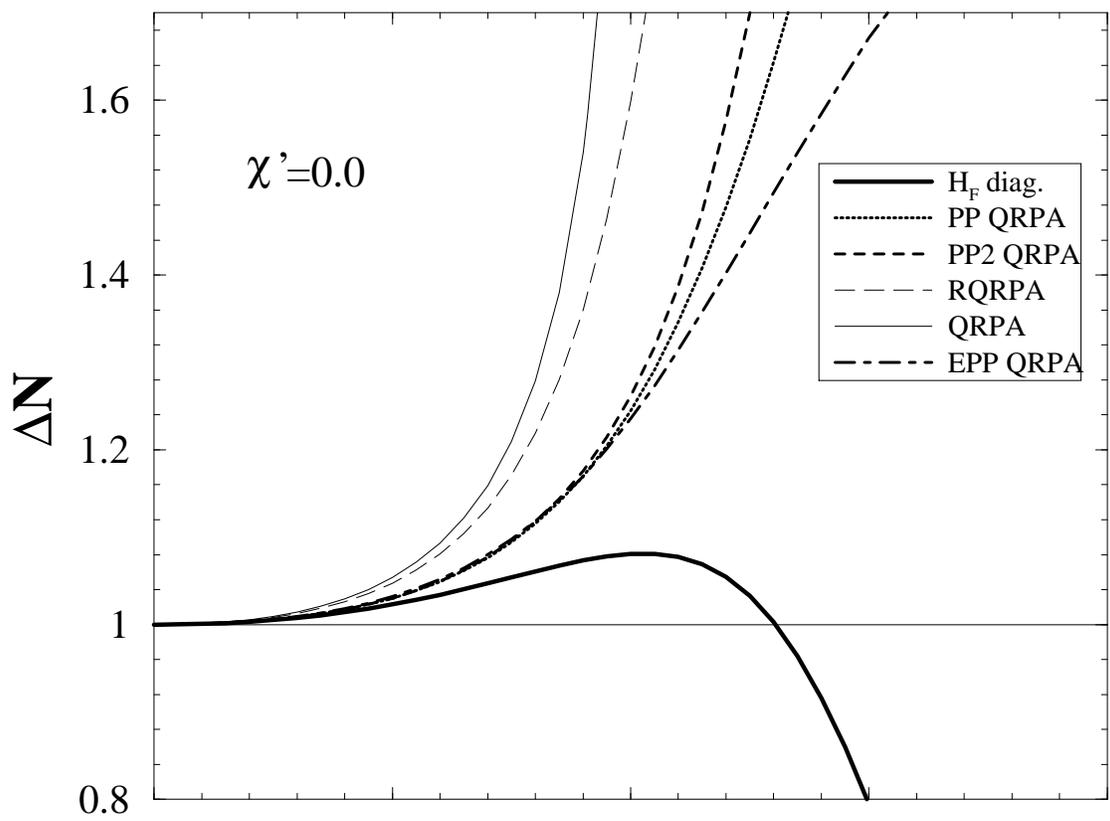
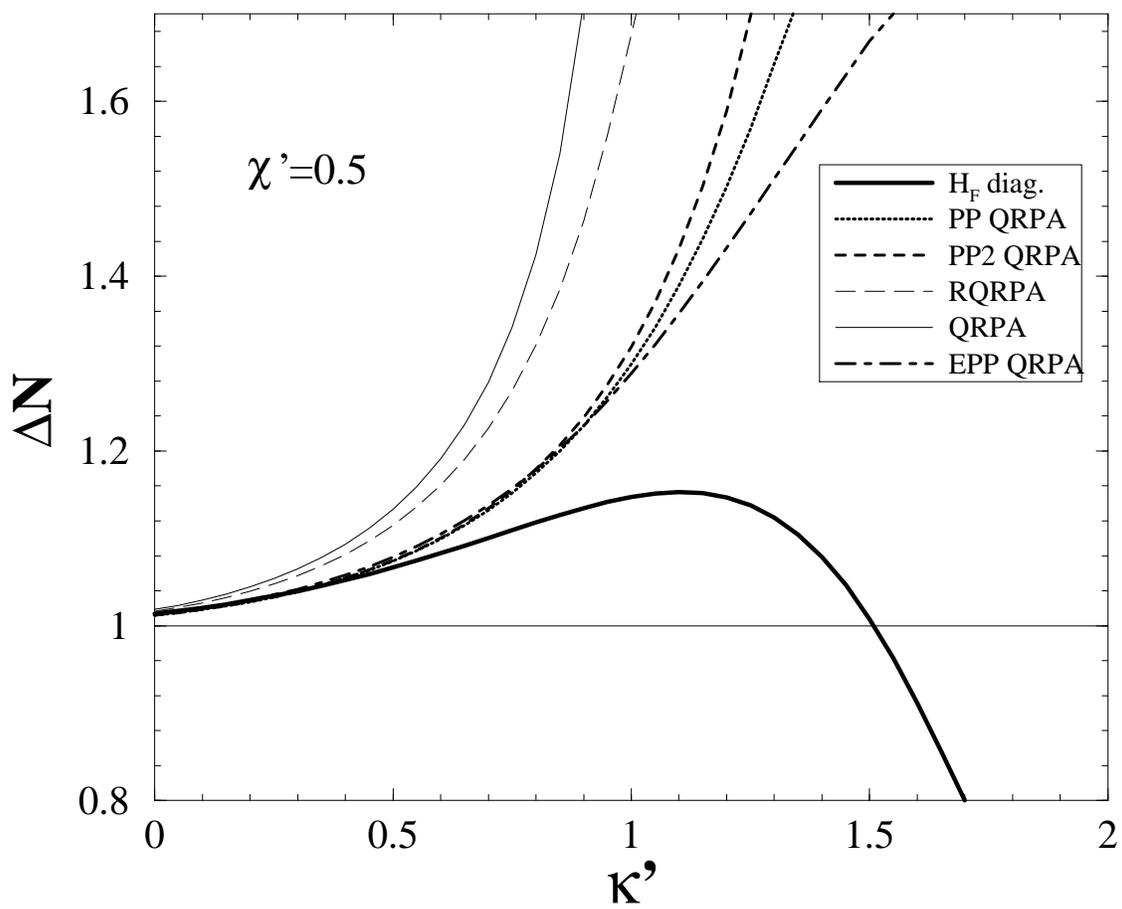
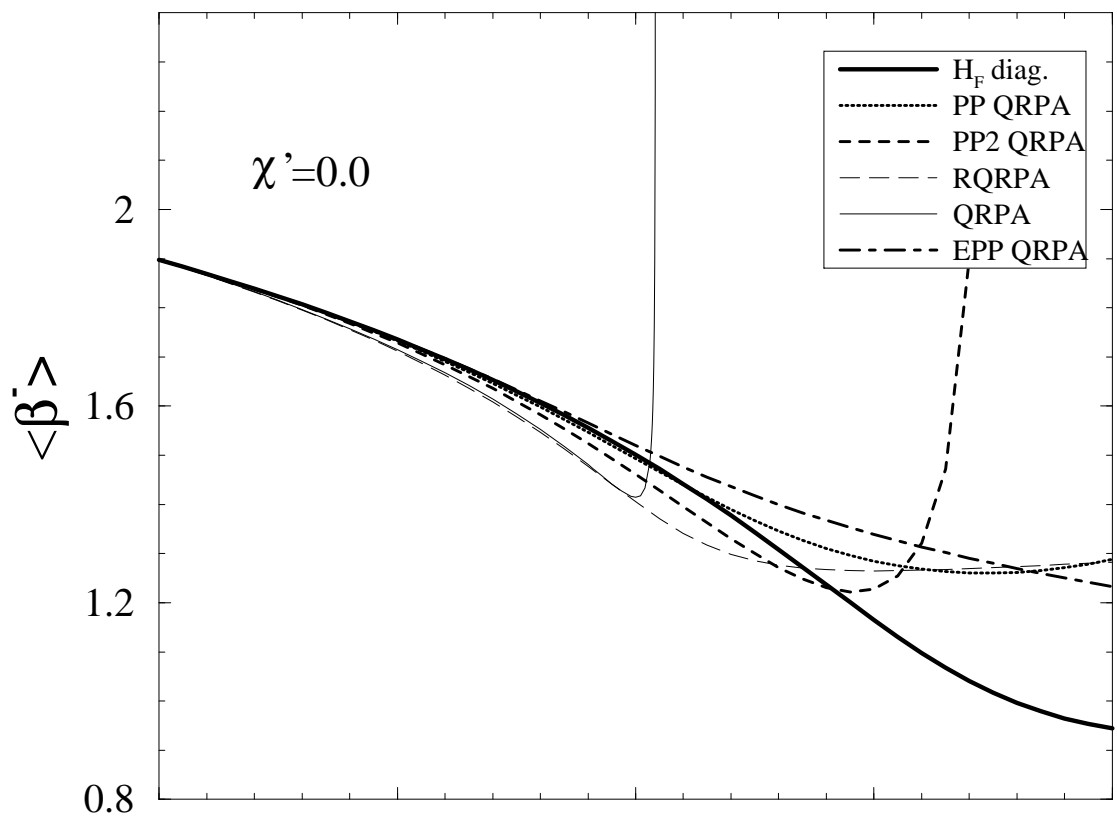
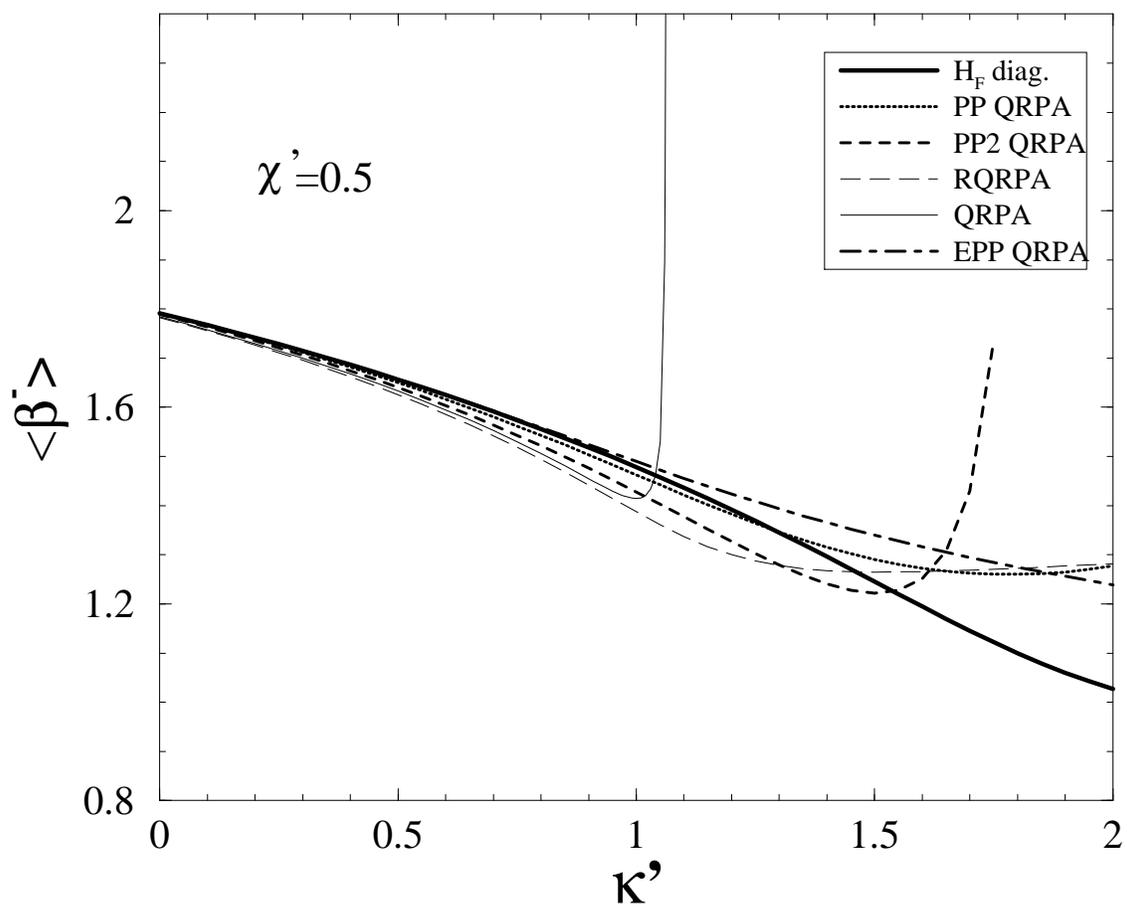
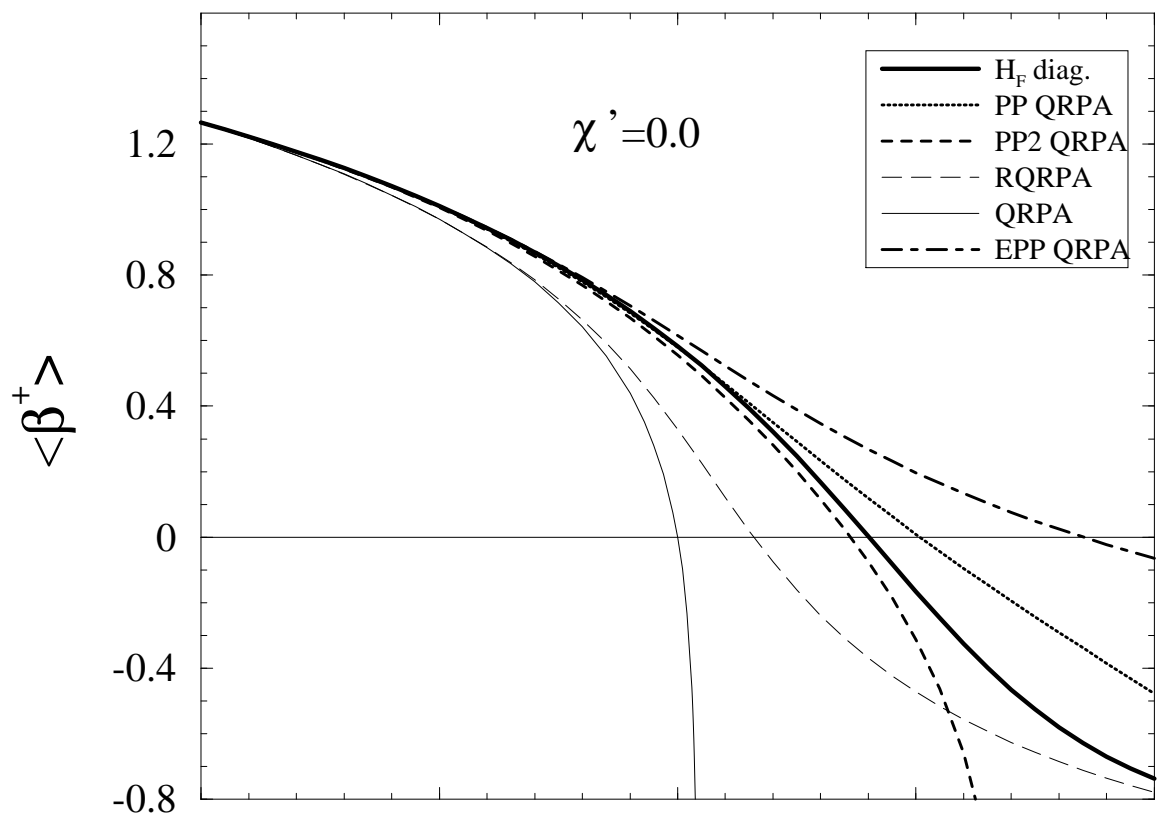
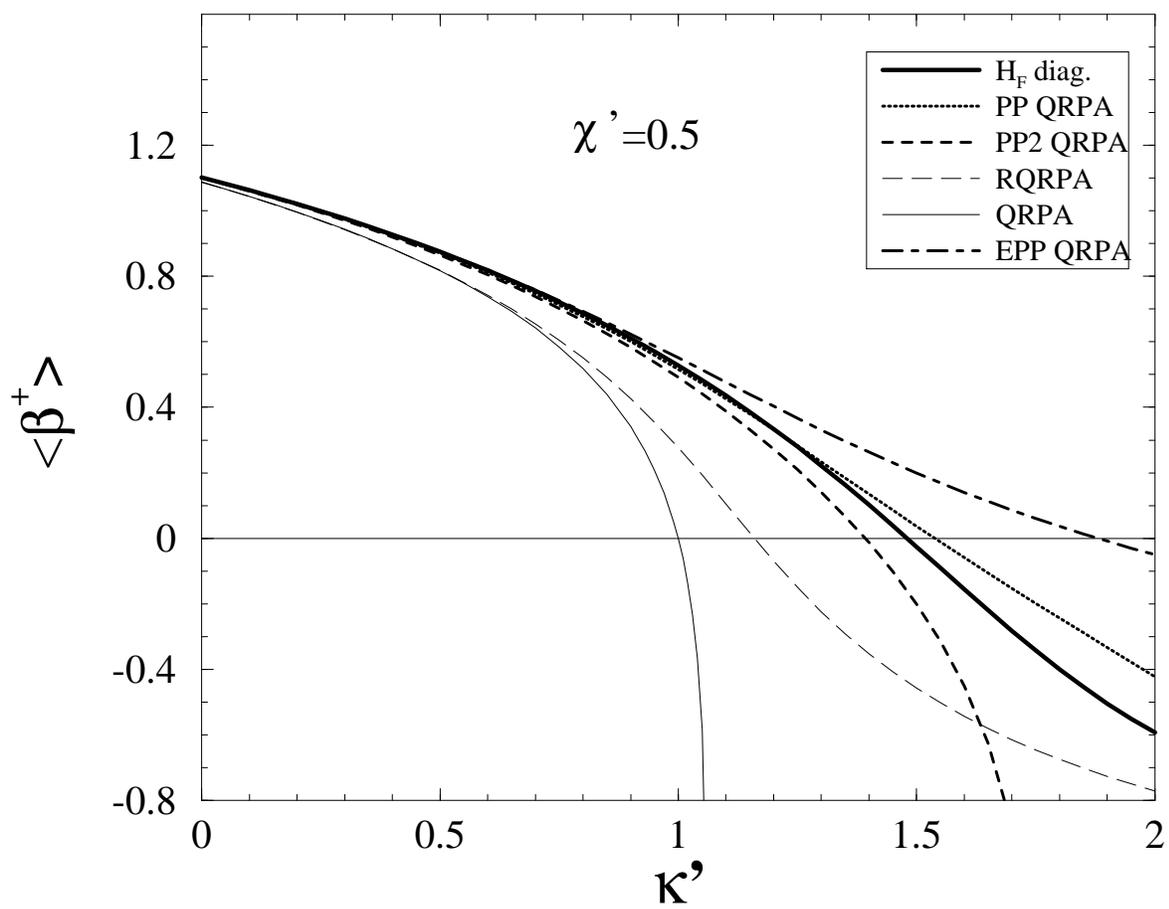
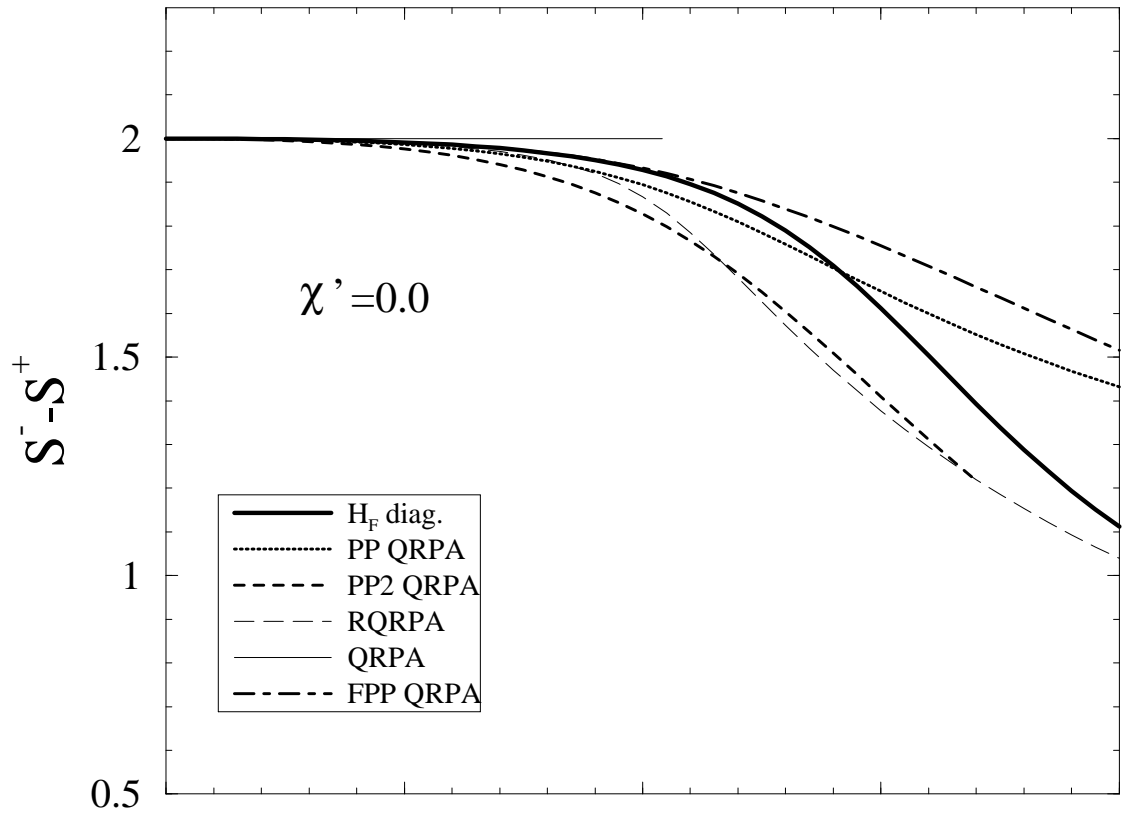
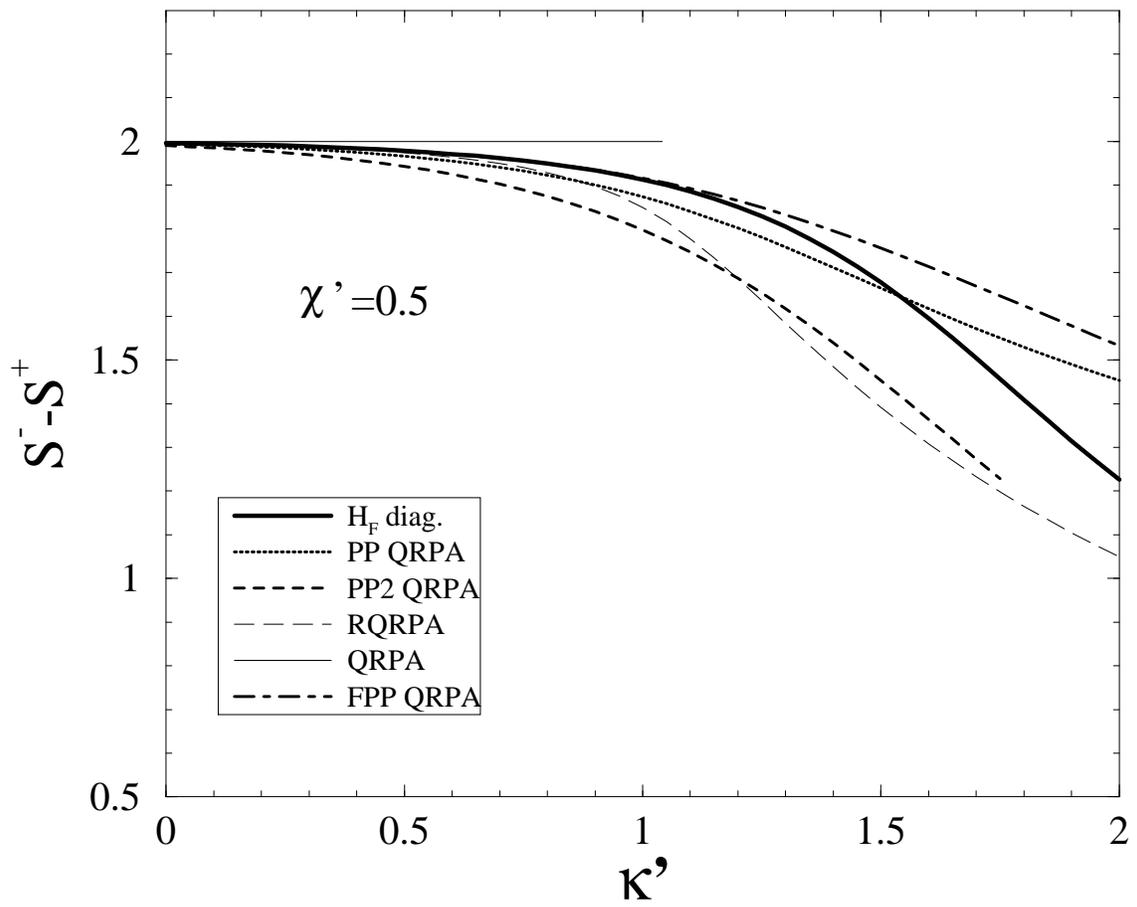
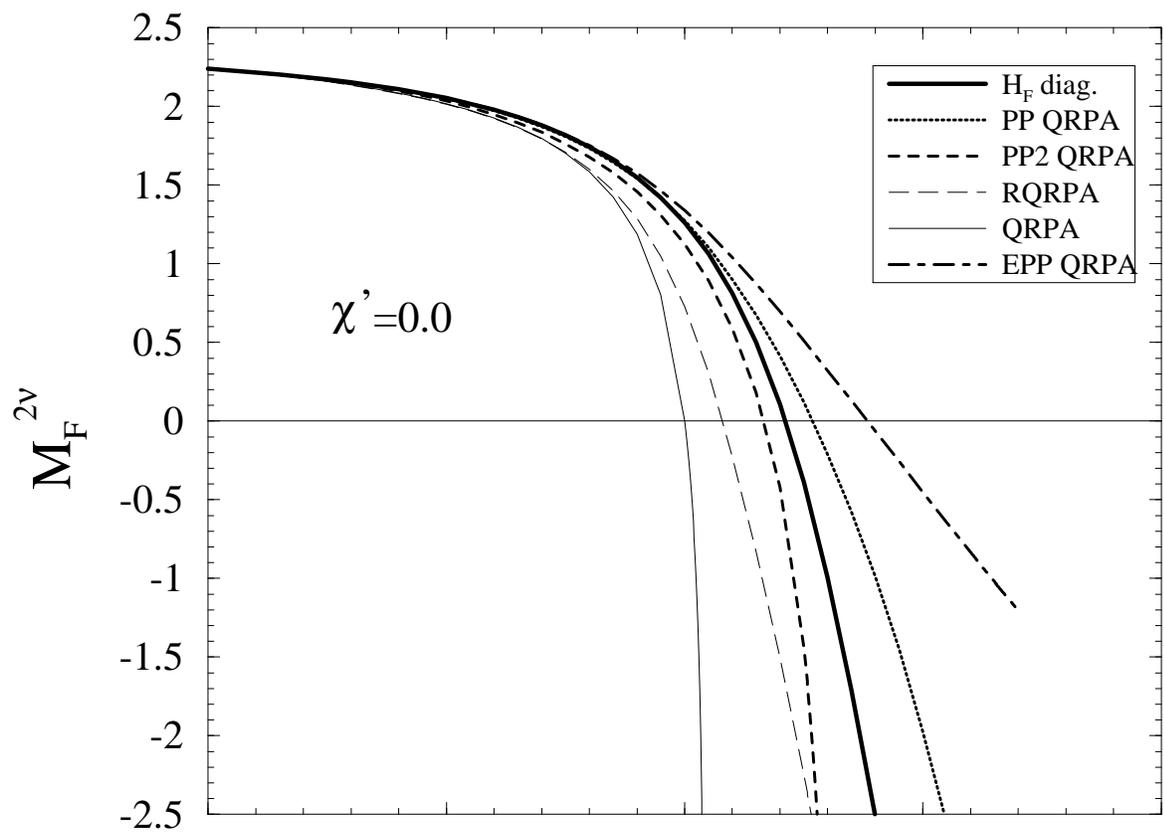
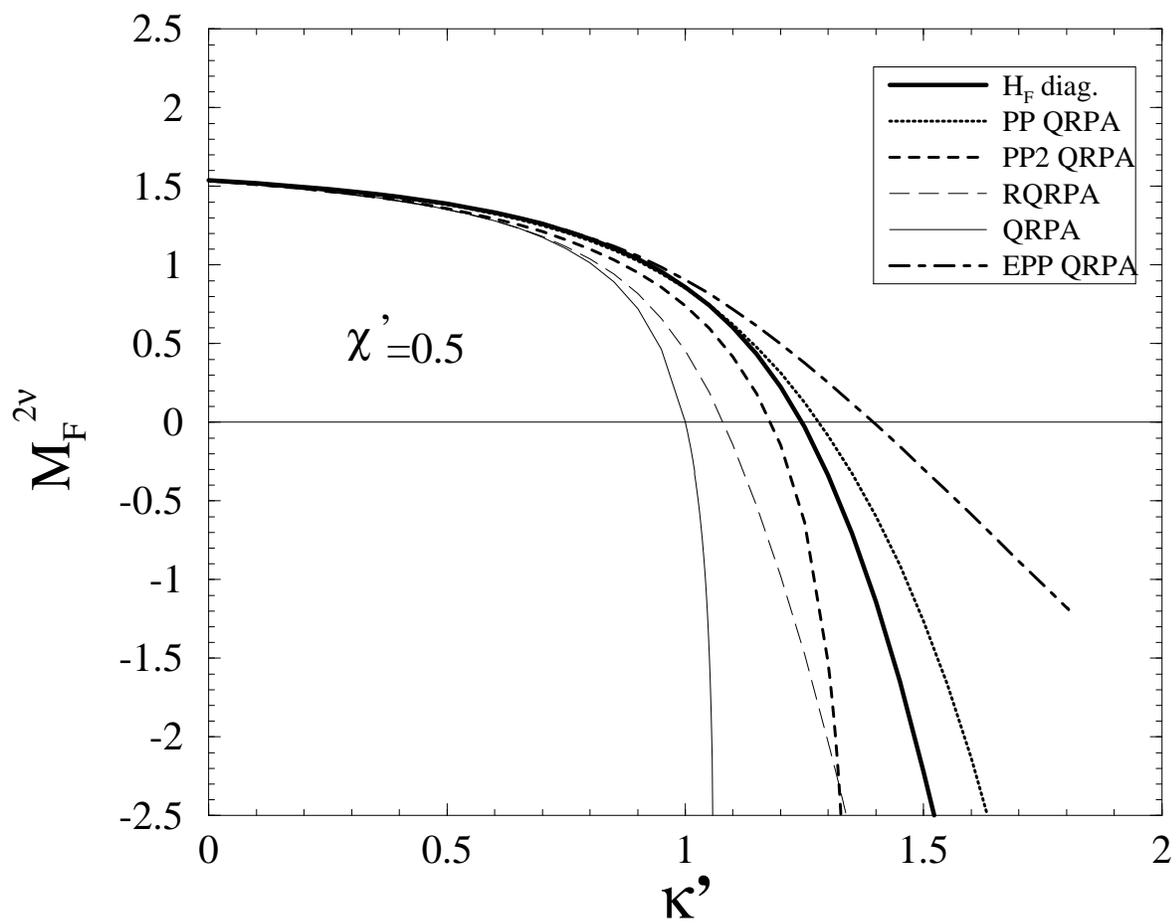
\end{document}